\documentclass[aps,prl,preprint,groupedaddress,showpacs]{revtex4} 
\usepackage{graphicx}
\usepackage{amssymb}

\begin{document}


\title{On the Switching between Negative and Positive \\Thermal Expansion in Framework Structures
}



\author{Andrea Sanson}
\email[E-mail address: ]{andrea.sanson@unipd.it}
\affiliation{Department of Physics and Astronomy - University of Padova, Padova (Italy)}



\date{\today}

\begin{abstract}

Thermal expansion is a problem in many technological applications and its control represents a challenge in materials design.
In this Letter, after studying the correlation between thermal expansion, cubic-to-rhombohedral transition and lattice parameter of ReO$_{3}$-type structures,
I show how in the different methods for controlling thermal expansion the key element is actually the lattice parameter. By varying the
lattice parameter through external pressure, chemical modifications or other methods, the single-well potential energy of the octahedral rotation, responsible for Negative Thermal Expansion (NTE), can be turned into a quartic anharmonic potential or into a double-well potential, thus enhancing or suppressing the NTE, respectively. This result can be extended to other framework-structures and should be taken into account to overcome the challenge of controlling thermal expansion.

\end{abstract}



\maketitle

The vast majority of materials expand on heating, while rare exceptions contract with increasing temperature. This fascinating property, known as negative thermal expansion (NTE), is very interesting both from a scientific and technological perspective. As a matter of fact, thermal expansion represents a problem for many materials and engineering applications, because when two coupled-materials expand differently when heated, this can lead to thermal shock breakage and failures of the system. For this reason, controlling thermal expansion represents a challenge for material design, and NTE is the starting point to develop materials with controlled thermal expansion \cite{Takenaka12,Chen15}.

Different methods are under investigation to control the thermal expansion. The application of external pressure represents a possible physical route to this goal. Indeed, on the basis of thermodynamics considerations, it is expected that the positive thermal expansion is diminished by the effect of pressure, while the negative thermal expansion is enhanced under pressure \cite{Dove16}. This behavior was experimentally confirmed, for example, in zinc cyanide \cite{Chapman07}. However, a few years ago, it was discovered that pressure induces switching between thermal contraction and expansion in ferroelectric PbTiO$_{3}$ \cite{Zhu14}, and similar reverse behavior was observed in other compounds \cite{Morelock13,Gallington17,Araujo18}.

Another promising route to control the thermal expansion is represented by chemical modification \cite{Chen15,Senn16}. The most famous example
is probably the case of metal fluorides with ReO$_{3}$-type structure, where the thermal expansion can be tuned from negative to zero to positive by acting on the chemical composition \cite{JChen17,CYang18}. Very interesting is the case of the cubic MZrF$_{6}$ series (M = Ca, Mn, Fe, Co, Ni, Zn), where the thermal expansion changes from about -6.7 to +18.2 $\times$ 10$^{-6}$ K$^{-1}$ \cite{Hu16}, and where CaZrF$_{6}$ displays a large NTE over a wide temperature range, much stronger than the most famous ZrW$_{2}$O$_{8}$ and other corner-sharing framework structures \cite{Hancock15}. Other interesting examples are given by Sc$_{1-x}$M$_{x}$F$_{3}$ (M = Y, Ti, Al, Ga, Fe) solid solutions \cite{Morelock13,Morelock14,Morelock15,Hu14}, where the precursor is scandium fluoride (ScF$_{3}$), another popular NTE material \cite{Greve10}. Recently, nano-size effects have been exploited to suppress the NTE of ScF$_{3}$ \cite{Hu18}.
\begin{figure}
\includegraphics[height=4cm,width=8cm]{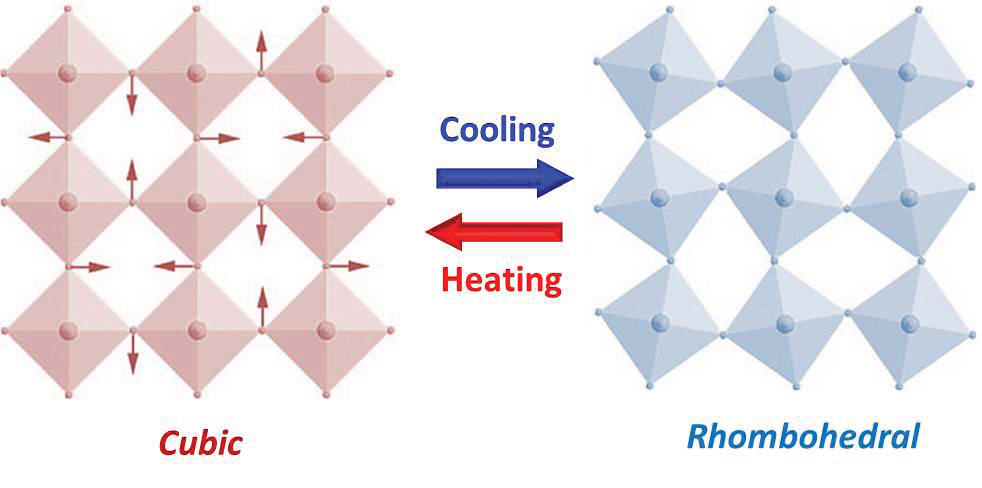}
\caption{Sketch of the cubic-rhombohedral structural phase-transition, which involves rigid rotations of the octahedra, in their cubic (left) and rhombohedral (right) forms. From the vibrational point of view, the rigid rotations of octahedra correspond to the triple-degenerate vibrational mode with F$_{1g}$ symmetry.} \label{f.1}
\end{figure}
\begin{figure*}
\begin{center}
\includegraphics[height=5.1cm,width=18cm]{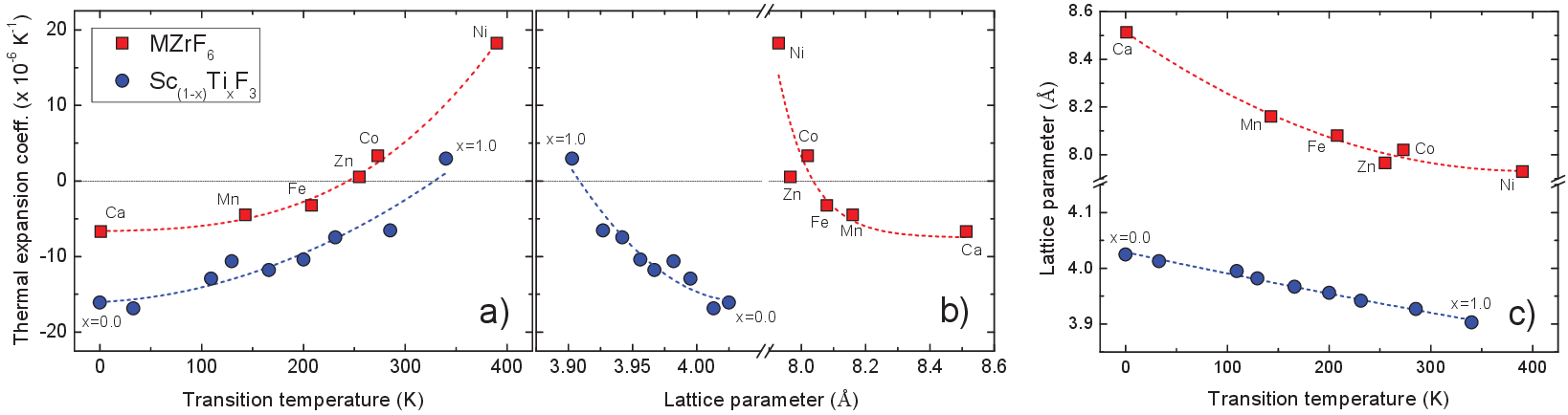}
\caption{Thermal expansion coefficient of MZrF$_{6}$ series (red squares) and Sc$_{1-x}$Ti$_{x}$F$_{3}$ compounds (blue circles) plotted as a function of the cubic-to-rhombohedral phase-transition temperature (panel (a)) and as a function of the lattice parameter (panel (b)). Panel (c) shows the  relationship between phase-transition temperature and lattice parameter. The dashed lines are a guide to the eyes. The experimental data were taken from average of Refs. \cite{Reinen78,Rodriguez90,Hu16,Hancock15,Morelock14}. For the lattice parameters, I used the values just before the cubic-to-rhombohedral transition.} \label{f.2}
\end{center}
\end{figure*}

The aim of this Letter is to understand how the pressure, the chemical modification or also other methods like nano-size effects, can act to tune the thermal expansion, shedding light on the switching between negative and positive thermal expansion and viceversa. I have adopted as reference for our study the MZrF$_{6}$ series. This series,
with the exception of CaZrF$_{6}$, show a cubic to rhombohedral structural phase transition which occurs with decreasing temperature \cite{Reinen78,Rodriguez90}. The same behavior is displayed by other metal fluorides with ReO$_{3}$-type structure \cite{Daniel90,Kennedy02}, including the Sc$_{1-x}$M$_{x}$F$_{3}$ solid solutions \cite{Morelock13,Morelock14,Morelock15,Hu14}. This transition, which leads to symmetry lowering and is frequently observed also under pressure \cite{Morelock15,Greve10}, can be visualized as rotation of octahedra about the crystallographic 3-fold axis (Fig. \ref{f.1}). Because rigid rotations of octahedra can be viewed as Rigid Unit Modes (RUM), which gives explanation for NTE \cite{Dove16,Tao03}, we can infer that the cubic to rhombohedral phase transition is in some way directly connected with the NTE behavior. On the other hand, it is well-known that RUMs softening in NTE framework structures is at the origin of structural phase transitions \cite{Dove96}.

Based on what we have just stated, in panel (a) of Fig. \ref{f.2} I have plotted the thermal expansion coefficient against the cubic-to-rhombohedral phase-transition temperature for the the MZrF$_{6}$ series. For M$=$Ca, I have assumed the transition temperature at 0 K, since no transition was experimentally observed in CaZrF$_{6}$, at least down to 10 K \cite{Hancock15}. It is interesting to observed the strict relationship between thermal expansion behavior and phase-transition temperature. The same relationship is present, for example, in Sc$_{1-x}$Ti$_{x}$F$_{3}$ compounds (blue circles in panel (a) of Fig. \ref{f.2}), where experimental data are available from Ref. \cite{Morelock14} (also here, for $x=0.0$, I have assumed the transition temperature at 0 K since no transition was observed in ScF$_{3}$ at least down to 10 K \cite{Greve10}).

Another very interesting relationship can be found between thermal expansion coefficient and lattice parameter in the cubic phase (panel (b) of Fig. \ref{f.2}), as well as between lattice parameter and phase-transition temperature (panel (c)): the NTE is progressively enhanced by increasing the lattice parameter, as well as the transition temperature increases as the lattice parameter decreases. From Fig. \ref{f.2}, it is evident that thermal expansion, cubic-to-rhombohedral transition and lattice parameter are directly correlated to each other, and the lattice parameter plays a fundamental role in the thermal expansion and in the phase-transition behavior. Hence, to get a deeper understanding of the origin of this close correlation, first-principles calculations based on density functional theory have been performed, focusing the attention
on the triple-degenerate vibrational mode with F$_{1g}$ symmetry of the MZrF$_{6}$ series. This vibrational mode, corresponding to rigid rotation
of MF$_{6}$ and ZrF$_{6}$ octahedra (Fig. \ref{f.1}) around M and Zr \cite{Sanson16}, respectively, plays a key role.
Indeed, it can be identified as a Rigid Unit Mode strongly contributing to NTE \cite{Sanson16},
and it can be strongly connected to the cubic-to-rhombohedral phase transition, which also involves rigid rotation of octahedra.

The CRYSTAL-14 package was used in this calculation, a periodic \emph{ab initio} program which uses a Gaussian-type
basis set to represent the crystalline orbitals \cite{Dovesi14}. For our purposes, I have chosen to study CaZrF$_{6}$, the one with the largest NTE and which shows no transition,
and CoZrF$_{6}$, which exhibits positive thermal expansion and cubic-to-rhombohedral transition at about room temperature (Fig. \ref{f.2}).
All-electron basis sets were employed, consisting of (9s)-(7631sp)-(621d) for zirconium atoms \cite{Valenzano11}, (7s)-(311sp) for fluorine \cite{Nada93}, (8s)-(6511sp)-(3d) for calcium \cite{Catti91}, and (8s)-(6411sp)-(41d) for cobalt \cite{Ruiz03}. Pure density functional theory calculations have been performed
using von Barth-Hedin \cite{vonBarth72} exchange and correlation functionals, which reproduce very well the experimental vibrational frequencies of the Raman active modes of CaZrF$_{6}$ \cite{Sanson16}. The Brillouin zone was sampled with a 12$\times$12$\times$12 k-mesh of Monkhorst-Park scheme \cite{Monkhorst76}, while the truncation criteria for bi-electronic integrals (Coulomb and exchange series), controlled by five parameters, was set to default values (7 7 7 7 14). To ensure good convergence,
the self-consistent-field convergence threshold on total energy was set to 10$^{-8}$ Hartree for initial geometry optimization,
to 10$^{-10}$ Hartree for frequency calculation and subsequent energy scan. More details on the computational aspects can be found in Refs. \cite{Dovesi14,Pascale04}.

The geometry optimization for CaZrF$_{6}$ gives a lattice parameter of about 8.4260 {\AA}
with fluorine position $x$=0.23889, while that for CoZrF$_{6}$ leads to a lattice parameter of about 7.8247 {\AA}
and fluorine position $x$=0.25695. These lattice parameters are in good agreement with the experimental data (Fig. \ref{f.2}),
whose difference is about 1.0 \% and 2.4 \%, respectively. However, when we calculate the vibrational frequencies at the
$\Gamma$-point of the Brillouin zone, we find that the vibrational mode with F$_{1g}$ symmetry, corresponding to rigid rotation
of MF$_{6}$ and ZrF$_{6}$ octahedra, has a negative frequency (about -1.09 THz) in the case of CoZrF$_{6}$. As expected, this means that
CoZrF$_{6}$ at 0 K is instable in the cubic form, unlike instead of CaZrF$_{6}$.

\begin{figure}
\begin{center}
\includegraphics[height=5.7cm,width=7cm]{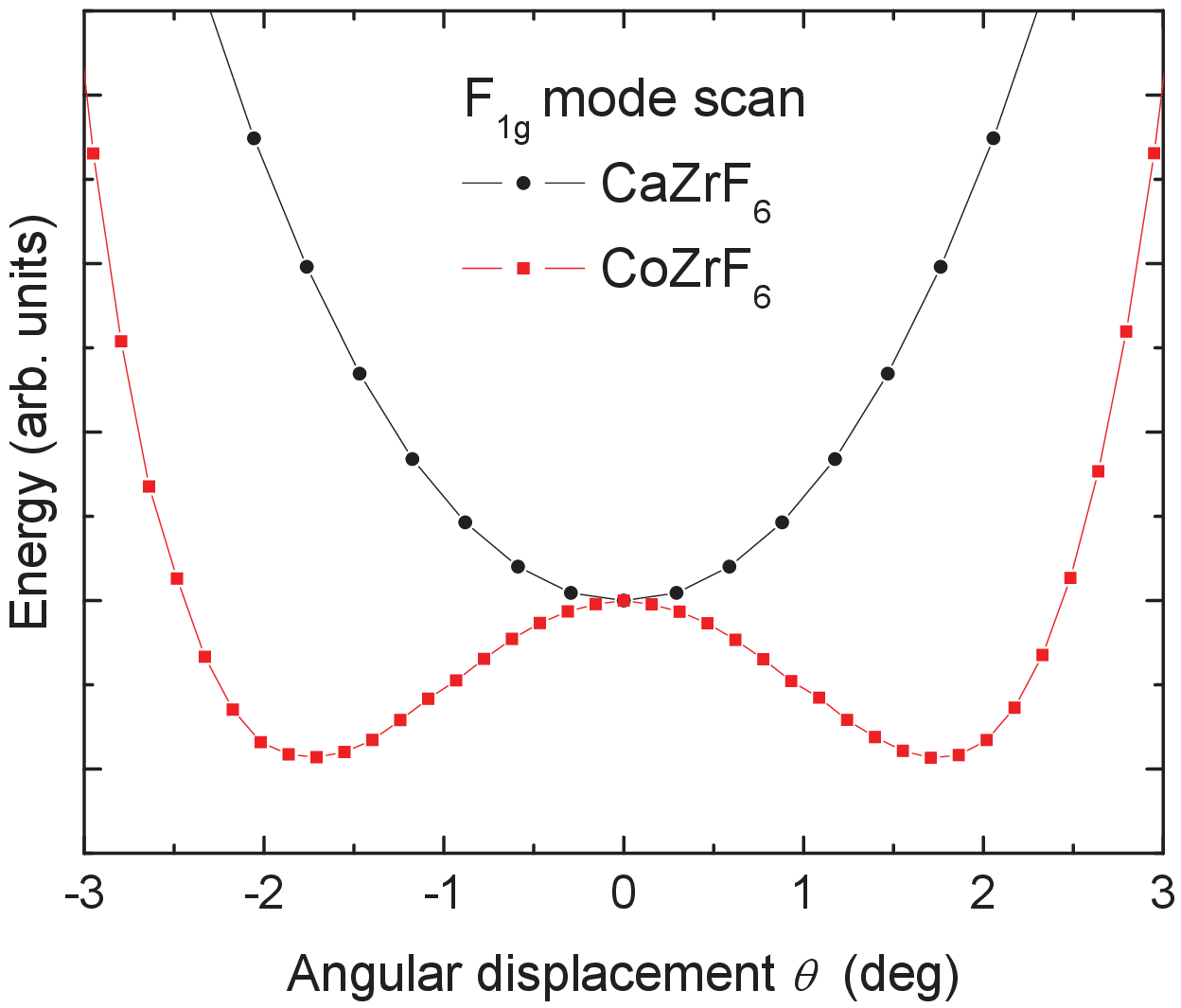}
\caption{Energy scan of the F$_{1g}$ vibrational mode, corresponding to rigid rotation of the ZrF$_{6}$ or MF$_{6}$ octahedra, in CaZrF$_{6}$ (black circles) and CoZrF$_{6}$ (red squares). The double-well potential found in CoZrF$_{6}$ is consistent with the cubic-to-rhombohedral transition observed experimentally at low temperature.} \label{f.3}
\end{center}
\end{figure}

By performing an energy scan of the F$_{1g}$ mode, i.e, the energy study as a function of angular rotation $\theta$ of ZrF$_{6}$ (or MF$_{6}$) octahedra \cite{note}, we can observe the existence of a double-well potential in the case of CoZrF$_{6}$ (Fig. \ref{f.3}), where the two minima correspond to equilibrium positions. Therefore, at low temperature, CoF$_{6}$ and ZrF$_{6}$ octahedra are rotated with respect to the cubic form, consistently with the rhombohedral form observed experimentally at low temperature (Fig. \ref{f.1}). In contrast, CaZrF$_{6}$ displays a single-well potential energy (Fig. \ref{f.3}), in agreement with the experimental absence of the cubic-to-rhombohedral transition at low temperature.
\begin{figure}
\begin{center}
\includegraphics[height=5.7cm,width=7.2cm]{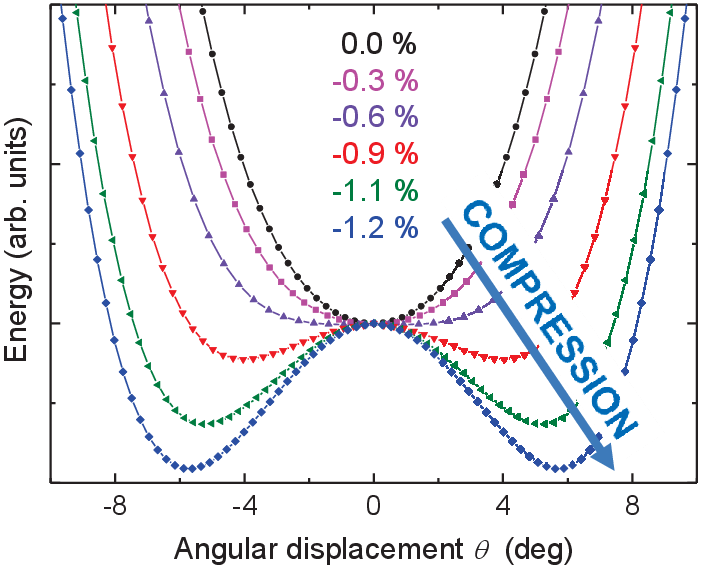}
\caption{Energy scan of the F$_{1g}$ vibrational mode in CaZrF$_{6}$ under compression, corresponding to a gradual reduction of the lattice parameter up to -1.2 \%: the single-well potential firstly turns into a quartic anharmonic potential, then it turns into a double-well potential, progressively deeper and with increasingly distant minima positions.} \label{f.4}
\end{center}
\end{figure}

According to Fig. \ref{f.2}, the lattice parameter should play a key role in the phase-transition and thermal expansion behavior. As a consequence,
by applying an external pressure (up to about 2.8 GPa), I have progressively reduced the lattice parameter of CaZrF$_{6}$ (up to -1.2 \%) and performed an energy scan of the F$_{1g}$ vibrational mode. The result is reported in Fig. 4. It is very interesting to observe that reducing the lattice parameter, the single-well potential firstly turns into a quartic anharmonic potential, then it turns into a double-well potential, progressively deeper and with increasingly distant minima positions (Fig. \ref{f.4}). This indicates that by reducing the lattice parameter, by external pressure, chemical modifications or other methods, we can create the conditions to generate the cubic-to-rhombohedral transition: the smaller the lattice parameter, the higher the phase-transition temperature (corresponding to the double-well potential depth) and the rhombohedral distortion. This explains the behavior shown in panel (c) of Fig. \ref{f.2}.
\begin{figure}
\includegraphics[height=13cm,width=7cm]{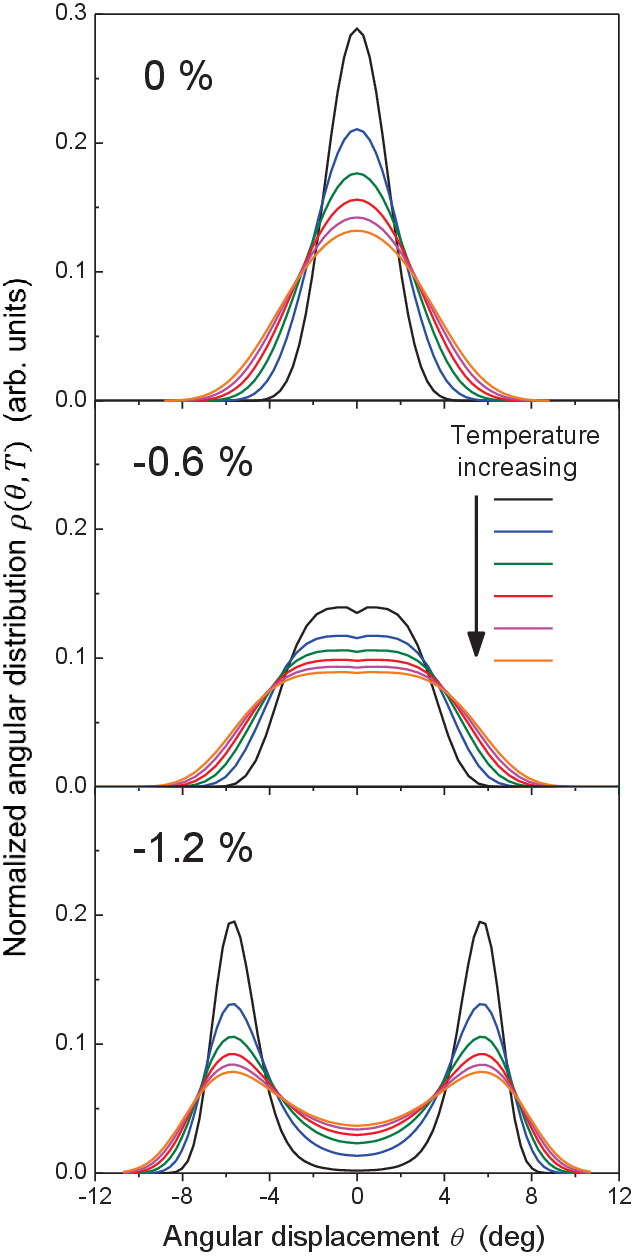}
\caption{Temperature evolution of the angular distribution $\rho(\theta,T)$ at selected lattice compressions: no compression (top panel), -0.6 \% compression (middle panel) and -1.2 \% compression (bottom panel). The angular distribution gradually widens with increasing lattice compression, up to split into two peaks at high compressions.} \label{f.5}
\end{figure}
\begin{figure}
\includegraphics[height=10cm,width=7.5cm]{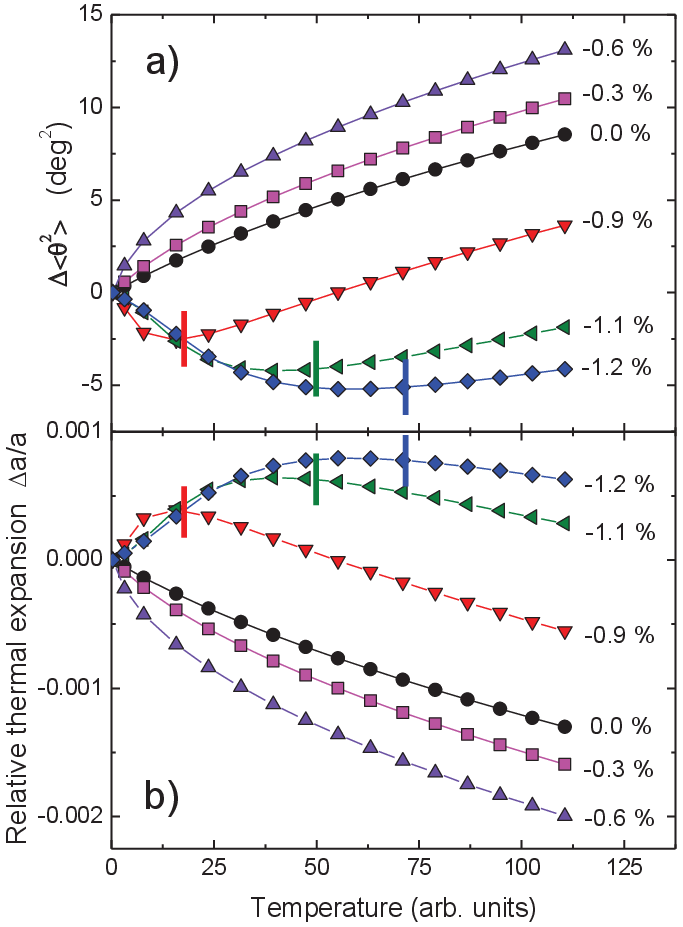}
\caption{Thermal evolution of the mean-square angular displacement $\langle\theta^{2}\rangle$ (panel (a)) and corresponding thermal expansion behavior (panel (b)) at different lattice compressions. The NTE is enhanced for lattice compressions up to -0.6 \%, to then switch into positive thermal expansion for higher compressions. The vertical bold-lines indicate the phase-transition temperature evident for the lattice compressions -0.9, -1.1 and -1.2 \%.} \label{f.6}
\end{figure}

At this point, what we are most interested in, is the thermal expansion behavior reported in Fig. \ref{f.2}. We know that the NTE is strongly related to RUMs \cite{Dove16,Tao03}, hence to the rotations of ZrF$_{6}$ and MF$_{6}$ octahedra. At a given temperature $T$, the lattice parameter $a(T)$ can be written as:
\begin{equation}\label{e.1}
a(T)=a_{0}\langle\cos\theta\rangle_{T}\simeq a_{0}\big(1-\langle\theta^{2}\rangle_{T}/2\big)
\end{equation}
where $a_{0}$ is the lattice parameter with no rotation, $\langle\theta^{2}\rangle_{T}$ is the mean-square angular displacement of the ZrF$_{6}$ (or MF$_{6}$) octahedra at temperature $T$. According to Eq. (\ref{e.1}), the distribution of the angular displacement, $\rho(\theta,T)$, plays a key role in the thermal expansion behavior. This distribution can be connected to the one-dimensional energy potential, $V(\theta)$, through the equation \cite{Cusack87}:
\begin{equation}\label{e.2}
\rho(\theta,T)=\exp[-V(\theta)/k_{B}T]\bigg\{\int\exp[-V(\theta)/k_{B}T]d\theta\bigg\}^{-1}
\end{equation}
where $k_{B}$ is the Boltzmann constant. Then using the energy potentials shown in Fig. \ref{f.4}, I have reconstructed the angular distributions $\rho(\theta,T)$ as a function of temperature for the different levels of lattice compression, as reported in Fig. \ref{f.5}. It can be observed that the angular distribution consists in one single distribution which gradually widens with increasing lattice compression, up to split into two peaks at high compressions (Fig. \ref{f.5}). This behavior is very important to explain the change of thermal expansion, including the switching between negative and positive thermal expansion and viceversa. In fact, by Eq. (\ref{e.1}), the relative thermal expansion is
\begin{equation}\label{e.3}
\frac{\Delta a}{a}\simeq-\Delta\langle\theta^{2}\rangle/2
\end{equation}
and depends on the "temperature evolution" of the mean-square angular displacement $\langle\theta^{2}\rangle$, which can be calculated at any temperature and compression by equation
\begin{equation}\label{e.4}
\langle\theta^{2}\rangle_{T}=\int\theta^{2}\rho(\theta,T)d\theta
\end{equation}
The panel (a) of Fig. \ref{f.6} shows the temperature evolution of the mean-square angular displacement $\langle\theta^{2}\rangle$ calculated at different lattice compressions, while the panel (b) shows the corresponding thermal expansion resulted from Eq. (\ref{e.3}). This figure allows us to derive the following outcomes:

\emph{i}) at low lattice compressions (here up to -0.6 \%) the NTE is enhanced. This because the single-well potential in which the octahedra rotate, turns into a quartic anharmonic potential (Fig. \ref{f.4}). The presence of a quartic potential increases the variation of $\langle\theta^{2}\rangle$ with temperature (Fig. \ref{f.6}a) and thus enhances the NTE (Fig. \ref{f.6}b). The origin of the strong NTE of ScF$_{3}$, explained in terms of quartic anharmonic potential \cite{ChenLi11}, can be placed in this context. In other words, ScF$_{3}$, unlike other metallic MF$_{3}$ fluorides, has a proper lattice parameter so to have a quartic anharmonic potential in which the ZrF$_{6}$ octahedra can rotate more freely, thus resulting in a very strong NTE;

\emph{ii}) at higher lattice compressions (here starting from -0.9 \%) the thermal expansion switches from negative to positive. This because the single-well potential in which the octahedra rotate, turns into a double-well potential (Fig. \ref{f.4}). The presence of a double-well potential reduces the variation of $\langle\theta^{2}\rangle$ with temperature (Fig. \ref{f.6}a) and thus inhibits the NTE (Fig. \ref{f.6}b). Note that above the phase-transition temperature, the thermal expansion returns negative but with lower magnitude.
The tuning of thermal expansion recently reached in ScF$_{3}$ nanoparticles and explained in terms of localized rhombohedral distortion \cite{Hu18}, falls in this
second case. By reducing the crystal size, from bulk to progressively smaller nanoparticles, the average lattice parameter becomes smaller and smaller \cite{Hu18}, therefore, according to Fig. \ref{f.4}, the single-well potential of the ScF$_{6}$ octahedra rotation turns into a double-well potential. This explains the observed rhombohedral distortion of ScF$_{3}$ in nano-form and the subsequent suppression of NTE.

In summary, in this Letter, thanks to the study of the relationship between thermal expansion, cubic-rhombohedral transition and lattice parameter of ReO$_{3}$-type structures, I have found out that the absolute value of the lattice parameter plays a key role in the control of thermal expansion. By varying the
lattice parameter through external pressure, chemical modifications or also other methods, the single-well potential in which the octahedra rotate can be turned into a quartic anharmonic potential or into a double-well potential, thus enhancing or inhibiting the NTE, respectively. This rule can be extended to other framework-structures and should be kept in mind to overcome the challenge of controlling thermal expansion.




\begin{references}


\bibitem{Takenaka12}
K. Takenaka, Negative thermal expansion materials: technological key for control of thermal expansion, Sci. Technol. Adv. Mater. \textbf{13}, 013001 (2012).

\bibitem{Chen15}
J. Chen, L. Hu, J. Deng, X. Xing, Negative Thermal Expansion in Functional Materials: Controllable Thermal Expansion by Chemical Modifications, Chem. Soc. Rev. \textbf{44}, 3522 (2015).

\bibitem{Dove16}
M. Dove and H. Fang, Negative thermal expansion and associated anomalous physical properties: review of the lattice dynamics theoretical foundation,
Rep. Prog. Phys. \textbf{79}, 066503 (2016).

\bibitem{Chapman07}
K.W. Chapman and P. J. Chupas, Pressure Enhancement of Negative Thermal Expansion Behavior and Induced Framework Softening in Zinc Cyanide,
J. Am. Chem. Soc. \textbf{129}, 10090 (2007).

\bibitem{Zhu14}
J. Zhu, J. Zhang, H. Xu, S. C. Vogel, C. Jin, J. Frantti, and Y. Zhao, Pressure-induced reversal between thermal contraction and expansion in ferroelectric PbTiO$_{3}$, Sci. Rep. \textbf{4}, 3700 (2014).

\bibitem{Morelock13}
C. R. Morelock, B. K. Greve, L. C. Gallington, K. W. Chapman,
and A. P. Wilkinson, Negative thermal expansion and compressibility of Sc$_{1-x}$Y$_{x}$F$_{3}$ ($x\leq$0.25),
J. Appl. Phys. \textbf{114}, 213501 (2013).

\bibitem{Gallington17}
L. C. Gallington, B. R. Hester, B. S. Kaplan, and A. P. Wilkinson, Pressure-dependence of the phase transitions and thermal expansion in
zirconium and hafnium pyrovanadate, J. Sol. St. Chem. \textbf{249}, 46 (2017).

\bibitem{Araujo18}
L. R. Araujo, L. C. Gallington, A. P. Wilkinson, and J. S.O. Evans, Phase behaviour, thermal expansion and compressibility of SnMo$_{2}$O$_{8}$,
J. Sol. St. Chem. \textbf{258}, 885 (2018).


\bibitem{Senn16}
M. S. Senn, C. A. Murray, X. Luo, L. Wang, F.-T. Huang, S.-W. Cheong, A. Bombardi, C. Ablitt, A. A. Mostofi, and N. C. Bristowe,
Symmetry Switching of Negative Thermal Expansion by Chemical Control, J. Am. Chem. Soc. \textbf{138}, 5479 (2016).

\bibitem{JChen17}
J. Chen, Q. Gao, A. Sanson, X. Jiang, Q. Huang, A. Carnera, C. Guglieri Rodriguez, L. Olivi, L. Wang, L. Hu, K. Lin,
Y. Ren, L. Gu, Z. Lin, C. Wang, J. Deng, J. P. Attfield, X. Xing,
Tunable thermal expansion in framework materials through redox intercalation, Nature Comm. \textbf{8}, 14441 (2017).

\bibitem{CYang18}
C. Yang, Y. Zhang, J. Bai, B. Qu, P. Tong, M. Wang, J. Lin, R. Zhang, H. Tong, Y. Wu, W. Songa, Y. Sun,
Crossover of thermal expansion from positive to negative by removing the excess fluorines in cubic ReO$_{3}$-type TiZrF$_{7-x}$,
J. Mater. Chem. C \textbf{6}, 5148 (2018).

\bibitem{Hu16}
L. Hu, J. Chen, J. Xu, N. Wang, et al., Atomic Linkage Flexibility Tuned Isotropic Negative, Zero, and Positive Thermal Expansion in MZrF$_{6}$ (M = Ca, Mn, Fe, Co, Ni, and Zn), J. Am. Chem. Soc. \textbf{138}, 14530 (2016).

\bibitem{Hancock15}
J.C. Hancock, K.W. Chapman, G.J. Halder, C.R. Morelock, B.S. Kaplan, L.C. Gallington, A. Bongiorno, C. Han, S. Zhou, A.P. Wilkinson, Large Negative Thermal Expansion and Anomalous Behavior on Compression in Cubic ReO$_{3}$-Type A$^{II}$B$^{IV}$F$_{6}$: CaZrF$_{6}$ and CaHfF$_{6}$, Chem. Mater. \textbf{27}, 3912 (2015).

\bibitem{Morelock14}
C. R. Morelock,L. C. Gallington,and A. P. Wilkinson, Evolution of Negative Thermal Expansion and Phase Transitions in
Sc$_{1-x}$Ti$_{x}$F$_{3}$, Chem. Mater. \textbf{26}, 1936 (2014).

\bibitem{Morelock15}
C. R. Morelock, L. C. Gallington, and A. P. Wilkinson, Solid solubility, phase-transitions, thermal expansion,
and compressibility in Sc$_{1-x}$Al$_{x}$F$_{3}$, J. Sol. St. Chem. \textbf{222}, 96 (2015).

\bibitem{Hu14}
L. Hu, J. Chen, L. Fan, Y. Ren, Y. Rong, Z. Pan, J. Deng, R. Yu, and X. Xing, Zero Thermal Expansion and Ferromagnetism in Cubic Sc$_{1-x}$M$_{x}$F$_{3}$
(M = Ga, Fe) over a Wide Temperature Range, J. Am. Chem. Soc. \textbf{136}, 13566 (2014).

\bibitem{Greve10}
B. K. Greve, K. L. Martin, P. L. Lee, P. J. Chupas, K. W. Chapman, and A. P. Wilkinson,
Pronounced Negative Thermal Expansion from a Simple Structure: Cubic ScF$_{3}$, J. Am. Chem. Soc. \textbf{132}, 15496 (2010).

\bibitem{Hu18}
L. Hu, F. Qin, A. Sanson, L.-F. Huang, Z. Pan, Q. Li, Q. Sun, L. Wang, F. Guo, U. Aydemir, Y. Ren, C. Sun, J. Deng, G. Aquilanti, J. M. Rondinelli, J. Chen, and X. Xing, Localized Symmetry Breaking for Tuning Thermal Expansion in ScF$_{3}$
Nanoscale Frameworks, J. Am. Chem. Soc. \textbf{140}, 4477 (2018).


\bibitem{Reinen78}
Von D. Reinen and F. Steffens, Struktur und Bindung in ubergangsmetall-Fluoriden M$^{II}$Me$^{IV}$F$_{6}$,
A. Phasenubergange, Z. anorg. allg. Chem. \textbf{441}, 63 (1978).

\bibitem{Rodriguez90}
V. Rodriguez, M. Couzi, A. Tressaud, J. Grannec, J. P. Chaminade, J. L. Soubeyroux, Structural phase transition in the ordered fluorides
M$^{II}$ZrF$_{6}$, (M$^{II}$ = Co, Zn): I. Structural study, J. Phys.: Condens. Matter \textbf{2}, 7373 (1990).

\bibitem{Daniel90}
P. Daniel, A. Bulou, M. Rousseau, J. Nouet, and M. Leblanc, Raman-scattering study of crystallized MF$_{3}$ compounds (M=Al,Cr,Ga,V,Fe,In): An approach to the short-range-order force constants, Phys. Rev. B \textbf{42}, 10545 (1990).

\bibitem{Kennedy02}
B.J. Kennedy and T. Vogt, Powder X-ray diffraction study of the rhombohedral to cubic phase transition in TiF$_{3}$
Mater. Res. Bull. \textbf{37}, 77 (2002).

\bibitem{Tao03}
J. Z. Tao and A. Sleight, The Role of Rigid Unit Modes in Negative Thermal Expansion,
J. Sol. St. Chem. \textbf{173}, 442 (2003).

\bibitem{Dove96}
K. D. Hammonds, M. Dove, A. P. Giddy, V. Heine, and B. Winkler, Rigid-unit phonon modes and structural phase transitions in framework silicates, Amer. Mineral. \textbf{81}, 1057 (1996).

\bibitem{Sanson16}
A. Sanson, M. Giarola, G. Mariotto, L. Hu, J. Chen, X. Xing,
Lattice dynamics and anharmonicity of CaZrF$_{6}$ from Raman spectroscopy and \emph{ab initio} calculations, Mater. Chem. Phys. \textbf{180}, 213 (2016).

\bibitem{Dovesi14}
R. Dovesi, R. Orlando, A. Erba, C.M. Zicovich-Wilson, B. Civalleri, S. Casassa,
L. Maschio, M. Ferrabone, M. De La Pierre, P. DArco, Y. Noel, M. Causa, M. Rerat,
B. Kirtman, CRYSTAL14: A program for the \emph{ab initio} investigation of crystalline solids, Int. J. Quantum Chem. \textbf{114}, 1287 (2014).

\bibitem{Valenzano11}
L. Valenzano, B. Civalleri, S. Chavan, S. Bordiga, M. Nilsen, S. Jakobsen, K. P. Lillerud, and C. Lamberti,
Disclosing the complex structure of UiO-66 MOF: a synergic combination of experiment and theory, Chem. Mater. \textbf{23}, 1700 (2011).

\bibitem{Nada93}
R. Nada, C.R.A. Catlow, C. Pisani, and R. Orlando, An \emph{ab-initio} Hartree-Fock perturbed-cluster study of neutral defects in LiF,
Model. Simul. Mater. Sci. Eng. \textbf{1}, 165 (1993).

\bibitem{Catti91}
M. Catti, R. Dovesi, A. Pavese, and V. R. Saunders, Elastic constants and electronic structure of fluorite (CaF$_{2}$): an \emph{ab initio} Hartree-Fock study,
J. Phys. Cond. Matter \textbf{3}, 4151 (1991).

\bibitem{Ruiz03}
E. Ruiz, M. Llunell, and P. Alemany, Calculation of exchange coupling constants in solid state transition
metal compounds using localized atomic orbital basis sets, J. Solid State Chem. \textbf{176}, 400 (2003).

\bibitem{vonBarth72}
U. von Barth and L. Hedin, A local exchange-correlation potential for the spin
polarized case. I, J. Phys. C: Solid State Phys. \textbf{5}, 1629 (1972).

\bibitem{Monkhorst76}
H. J. Monkhorst and J. D. Pack, Special points for Brillouin-zone integrations, Phys. Rev. B \textbf{13}, 5188 (1976).

\bibitem{Pascale04}
F. Pascale, C. M. Zicovich-Wilson, F. Lopez Gejo, B. Civalleri, R. Orlando, and R. Dovesi,
The calculation of vibrational frequencies of crystalline compounds
and its implementation in the CRYSTAL code, J. Comput. Chem. \textbf{25}, 888 (2004).

\bibitem{note}
Since the Zr-F and M-F bond length are similar, the angular displacement $\theta$ of the ZrF$_{6}$ and MF$_{6}$ octahedra will be very similar to each other.

\bibitem{Cusack87}
N. E. Cusack, \emph{The Physics of Structurally Disordered Matter}, Bristol: Adam Hilger (1987).

\bibitem{ChenLi11}
Chen W. Li, X. Tang, J. A. Mu\~{n}oz, J. B. Keith, S. J. Tracy, D. L. Abernathy, and B. Fultz,
Structural Relationship between Negative Thermal Expansion and Quartic Anharmonicity of Cubic ScF$_{3}$,
Phys. Rev. Lett. \textbf{107}, 195504 (2011).




\end{references}
\end{document}